\documentclass[pdflatex,sn-mathphys-num]{sn-jnl}
\usepackage{amsmath,amssymb,amsfonts}
\usepackage{mathtools} 
\usepackage{graphicx}
\hypersetup{colorlinks=true,pdfborder={0 0 0.1},linkcolor=orange!50!black,urlcolor=blue!30!black,citecolor=blue!60!black,
          pdftitle={Tensor product algorithms for inference of contact network from epidemiological data},
          pdfauthor={S. V. Dolgov, D. V. Savostyanov},
}
\usepackage[ruled]{algorithm}
\usepackage{algpseudocode}
\usepackage{xcolor}
\def\hat{\mathaccent"705E }
\def\eps{\varepsilon}
\def\geq{\geqslant} \def\leq{\leqslant}
\def\Real{\mathbb{R}} \def\Binary{\mathbb{B}} \def\Permutations{\mathfrak{S}}
\def\Uniform{\mathcal{U}}
\def\X{\mathbb{X}}            
\def\O{\mathcal{O}}
\def\inf{\text{(inf)}} \def\rec{\text{(rec)}}   

\def\phi{\varphi}
\def\Prob{\mathsf{P}} \def\Mean#1{\mathsf{E}[#1]}  \def\Like{\mathsf{L}} 

\def\diag{\mathop{\mathrm{diag}}\nolimits}
\def\cnd{\mathbin{|}}

\def\evec{\vec e} \def\svec{\vec s} \def\ivec{\vec{\text{\emph{\i}}}}   
\def\x{\mathrm{x}} \def\y{\mathrm{y}} \def\e{\mathrm{e}} \def\u{\mathrm{u}} \def\v{\mathrm{v}}  
\def\p{\mathbf{p}}  
\def\pcore{\mathrm{p}}  
\def\J{\mathbf{J}} \def\A{\mathbf{A}}  
\def\Id{\mathrm{Id}}
\def\zero{\color{black!40}{0}}
\def\Indicator#1{\mathbf{1}_{#1}}
\def\Nodes{\mathcal{V}} \def\Edges{\mathcal{E}}  
\def\Grid{\mathcal{G}} \def\Grids{\mathbb{G}}
\def\Data{\mathcal{X}}
\def\Infected{\mathcal{I}}

\def\Nssa{N_{\mathrm{SSA}}} \def\nssa{n_{\mathrm{SSA}}}
\def\neval{N_{\mathrm{eval}}}
\def\tmax{T} 
\def\tstep{\Delta t}
\def\Temp{\tau}
\begin{document}

\title[Tensor product algorithms for network inference]{Tensor product algorithms for  inference of contact network from epidemiological data\footnote{%
 This version of the article has been accepted for publication in \emph{BMC Bioinformatics}, after peer review, but is not the Version of Record and does not reflect post-acceptance improvements, or any corrections. 
 The Version of Record is available online at: \href{https://dx.doi.org/10.1186/s12859-024-05910-7}{doi:10.1186/s12859-024-05910-7}.}}

\author[1]{\fnm{Sergey} \sur{Dolgov}}\email{S.Dolgov@bath.ac.uk}
\equalcont{These authors contributed equally to this work.}

\author*[2]{\fnm{Dmitry} \sur{Savostyanov}}\email{D.Savostyanov@essex.ac.uk}
\equalcont{These authors contributed equally to this work.}

\affil[1]{\orgname{University of Bath}, \orgaddress{\street{Claverton Down}, \city{Bath}, \postcode{BA2 7AY}, \country{UK}}}
\affil*[2]{\orgname{University of Essex}, \orgaddress{\street{Wivenhoe Park}, \city{Colchester}, \postcode{CO4 3SQ}, \country{UK}}}

\abstract{
  We consider a problem of inferring contact network from nodal states observed during an epidemiological process.
  In a black--box Bayesian optimisation framework this problem reduces to a discrete likelihood optimisation over the set of possible networks.
  The cardinality of this set grows combinatorially with the number of network nodes, which makes this optimisation computationally challenging.
  For each network, its likelihood is the probability for the observed data to appear during the evolution of the epidemiological process on this network.
  This probability can be very small, particularly if the network is significantly different from the ground truth network,  from which the observed data actually appear.
  A commonly used stochastic simulation algorithm struggles to recover rare events and hence to estimate small probabilities and likelihoods.
  In this paper we replace the stochastic simulation with solving the chemical master equation for the probabilities of all network states.
  Since this equation also suffers from the curse of dimensionality, we apply tensor train approximations to overcome it and enable fast and accurate computations.
  Numerical simulations demonstrate efficient black--box Bayesian inference of the network.
}

\maketitle

\section{Introduction} \label{sec:intro}        

\subsection{Background}
The recent outbreak of COVID-19 and the public discussion that followed has led to better understanding of the central role that epidemiological models play in the decision making process and developing an informed response strategy.
The quality of the mathematical models used in this process is crucial, not only because it allows to accurately predict the spread of a disease in population, but also in order to increase public trust in research and the decisions that are based on it.
A large number of epidemic models used in education and research follow the Kermack--McKendrick compartmental SIR model~\cite{kmk-sir-1927}.
An implicit assumption of this model is that the susceptible, infected and recovered groups are well--mixed in the sense that every person in any group has the same probability to come in contact with anyone from another group, i.e. the contact network is homogeneous.
However, this assumption holds relatively poorly in communities, which poses a significant limitation to application of compartmental models.
More accurate models should include the information on the structure of contact network, leading to study of epidemics on networks~\cite{keeling-network-survey-2005,chen-stoch-sir-2005,youssef-network-sir-2011,mieghem-network-2015,kiss-network-2017}.

Predicting the evolution of epidemic on a given network is much more difficult than solving a homogeneous model.
Consider the case when each network node represents a single individual, who can be susceptible or infected.
The dynamics of the epidemic becomes stochastic and depends on the position of infected individuals in the network and the number of susceptible neighbours they are in contact with.
Hence, we need to model a probability distribution function of network states instead of states themselves.
This probability distribution function satisfies the chemical master equation (CME)~\cite{vankampen-stochastic-1981}, which is an ordinary differential equation on the probability values. 
However, the total number of network states (and hence the size of the CME) grows exponentially in the number of network nodes~\cite{chen-stoch-sir-2005,youssef-network-sir-2011}.
This makes the direct solution of the stochastic network models computationally intractable for large networks~\cite{kiss-network-2017}.

Perhaps the most traditional method for tackling the CME is the Stochastic Simulation Algorithm (SSA) \cite{gillespie-ssa-1976} and variants, which compute random walks over the network states, distributed according to the CME solution.
However, as a Monte Carlo method, the SSA is known to converge slowly, especially for rare events.
Alternative approaches include
 mean--field approximations~\cite{keeling-meanfield-1999,rand-meanfield-1999},
 effective degree models~\cite{gleeson-degree-2011,lindquist-degree-2011,kiss-degree-2012},
 and edge--based compartmental models~\cite{miller-edge-2012}, but these models are approximate and rely on truncation of the state space. This introduces a truncation error that is difficult to estimate and/or keep below a desired tolerance for a general network.
Other approaches include changing the original model into a surrogate model such as birth--death processes~\cite{kiss-infer-class-2020}, or using neural networks~\cite{Khammash-NN-CME-2021,Grima-CME-NN-2022}.
For solving the original model in a numerically controllable approximation framework, a new approach based on tensor product factorisations was recently proposed by authors in~\cite{ds-ttsir-2024}.
 
If the contact network is not known, we can attempt to solve an inverse problem, i.e. to infer the network from observations of disease data over time.
For $N$ network nodes, the number of possible networks grows exponentially in $N^2.$ 
Hence, for large $N,$ the problem complexity typically grows much faster than the information available, and network inference becomes a (very) under-determined problem~\cite{marchal-infer-review-2010}.
Network inference is therefore only solved directly for very small population sizes~\cite{roberts-infer-bayes-1999,lopez-sis-2015}, 
   or equivalently by assuming that the population consists of a few densely connected groups and estimating couplings between them~\cite{keeling-coupling-2002}.
Networks with mass-action kinetics can be inferred uniquely by observing transition rates at a simplex set of states \cite{Erban-infer-2021}, but the cardinality of this set is combinatorial in $N$.
To address the problem for larger $N,$ one can involve additional information about the network structure, such as
   degree distribution and sparsity~\cite{mukherjee-infer-priors-2008,di-infer-sparse-2015},
   community structure~\cite{peixoto-infer-community-2019},
   and/or assumptions on the underlying statistical distribution for the network and infer its parameters~\cite{hunter-bayesian-2011}.
   In some cases more complex than pairwise interactions improve the network reconstruction~\cite{landry-duality-2022}.
   Predictability of a stochastic process of observations and reconstructability of a network
   using their mutual information was considered in~\cite{murphy-duality-2024}.
   This information can be used to estimate the success of a network inference, before the full inference algorithm is applied.
   In contrast to stochastic inference, one can instead model a deterministic dynamics and minimise the observational error over the network parameters in the right-hand side of the dynamics~\cite{shandilya-infer-2011}.
Another approach is to 
  infer properties of the network rather than its exact structure~\cite{britton-infer-egocentric-2015},
  e.g. to find a class of network distributions, which the contact network most likely belongs to~\cite{kiss-infer-class-2020}.
Related work include inferring the origin of epidemic given the contact network~\cite{lokhov-infer-origin-2014}.
For a recent survey on network inference see~\cite{wolf-infer-survey-2018}.
Finally, most close to our work is the maximum-likelihood estimation
of the network link probabilities from binary time series~\cite{ma-infer-2018}.
However, the latter paper uses an expectation-maximisation algorithm assuming the Poisson distribution, while in this paper we rely on the full Bayesian formalism with likelihoods computed directly from the chemical master equation.

\subsection{Our contribution}
\begin{figure}[t]
 \begin{center}
   \includegraphics[width=.95\textwidth]{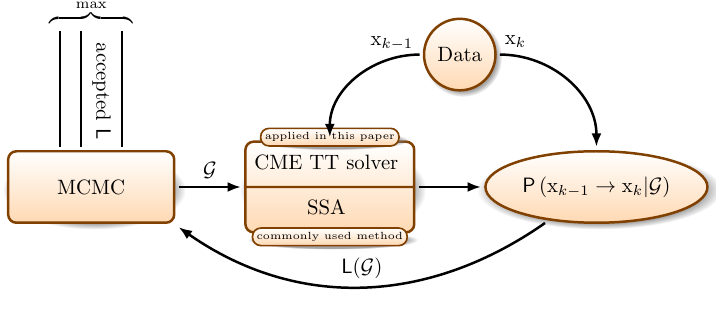}
 \end{center}
 \caption{Network inference workflow. 
 An MCMC algorithm samples proposal network configurations, $\Grid$. 
 The CME is solved on each time interval $[t_{k-1},t_k]$ in the observed data, starting from the state observation $X(t_{k-1})=\x_{k-1}$ and obtaining the TT approximation of the probability of observing the state $X(t_{k})=\x_{k}$ for the given network $\Grid.$
 The CME solver is applied in place of a more commonly used SSA method, that struggles to recover rare events.
 The probabilities for all data are multiplied to form the likelihood $\Like(\Grid)$, which is accepted or rejected in the MCMC. 
 Finally, the network with the maximum likelihood among the MCMC samples is inferred.
 } \label{fig:workflow}
\end{figure}
In this paper we investigate inference of the contact network from states of the network observed over time.
We use a Bayesian formulation to find the network with the maximum a posteriori (MAP) estimate.
However, we do not assume any prior knowledge on the structure of the contact network (uniform prior), and thus look for the maximum likelihood estimate (MLE),
\[
 \text{network}_{\text{opt}} = \arg\max_{\text{network}} \Prob( \text{data} \cnd \text{network} ),
\]
which we solve as black--box optimisation problem.
To summarise the above, the network inference problem is difficult due to the following reasons.
 \begin{enumerate}
  \item\label{item:inv}
   Large inverse problem. 
   The inverse problem is a high--dimensional discrete optimisation problem.
   Indeed, in an undirected network with $N$ nodes there are $\tfrac12 N (N-1)$ potential links, which are the optimisation variables, each of which can independently be in one of two states (on/off).
   In total, the search space which consists of $2^{N(N-1)/2}$ possible networks, hence the exhaustive search is computationally unfeasible.
   Since the optimisation variables are discrete, the gradient of the target function is unavailable, and hence we can't apply steepest descend or Newton--Raphson algorithms.
   Hence, to find the near--optimal solution, we need to explore the structure of the high--dimensional array with some (possibly heuristic) algorithm,  which may require a large number of target function evaluations.
  \item\label{item:fwd}
   Large forward problem.
   For each network in the search space, a single evaluation of the target function requires solving a forward problem, i.e. finding a probability of given data to be observed during the evolution of the disease on the current network.
   This is a Markov chain problem on the state space of accessible network states, which scales exponentially with the number of nodes, causing the currently available algorithms to struggle.
  \item\label{item:low} 
   Low contrast caused by insufficient data.
   We may observe conditional probabilities to be (almost) the same, 
   \(
   \Prob( \text{data} \cnd \text{network}_1 ) \approx \Prob( \text{data} \cnd \text{network}_2),
   \)
   for example if the two networks differ only by the links attached to the nodes, for which we do not have (enough) events in the observed dataset.
   In this case, the Bayesian optimisation won't be able to choose one particular network, as a large number of them are equally likely to produce the observed dataset.
   Any numerical approximation errors due to limitations of the forward problem solvers (see item~\ref{item:fwd}) add extra noise to the high-dimensional probability density function and further complicate the optimisation process (see item~\ref{item:inv}).
   Due to the probabilistic nature of the problem, the ground truth network $ \text{network}_\star, $ for which the observed dataset was generated, may differ from the optimal network recovered by the Bayesian inference method,
   \(
   \text{network}_\star \neq \text{network}_{\text{opt}}.
   \)
  \item\label{item:high} 
   High contrast caused by a large amount of data. 
   Although adding more data makes the ground truth network a unique global optimum for Bayesian optimisation, it also creates a large number of local optima.
   In a black--box optimisation setting, there is no prior information that could navigate the optimisation towards the global optimum, and the algorithm can be trapped in a local optimum for considerable time.
 \end{enumerate}

The existing literature often does not consider these issues separately, nor approach the problem directly.
It is typically stated that the network optimisation is impossible to solve, and alternative formulations are considered~\cite{marchal-infer-review-2010,kiss-network-2017,wolf-infer-survey-2018,kiss-infer-class-2020}.
In this paper we attempt to perform the black--box network inference directly following the Bayesian optimisation framework.
To tackle the forward problem, we solve the CME using tensor product factorisations~\cite{ds-ttsir-2024}.
A related work was recently proposed in tensor network community, but it is limited to linear one--dimensional chains~\cite{merbis-ttsis-2023}.
We demonstrate that the proposed method provides faster and more accurate solution to the forward problem compared to SSA, particularly when the network is far from optimal.

Next, we apply two Markov Chain Monte Carlo (MCMC) algorithms for black--box discrete high--dimensional optimisation and analyse results.
An overall workflow of this procedure is illustrated in Fig.~\ref{fig:workflow}.
By simulating three examples of networks, we show that by collecting sufficiently many data we can make the contrast high enough to infer the original ground truth network.

\subsection{Notation}
We use 
   calligraphic font for contact networks $\Grid,$
   blackboard font for sets $\Real,$ $\Grids,$ $\X,$
   sans-serif font for probability $\Prob$, expectation $\mathsf{E},$ variance $\mathsf{V},$ and likelihood $\Like.$
Indices $m,n$ and scalar values $x, p$ are shown in usual maths italics.
We use maths roman font for vectors, e.g. network states $\x$ or unit vectors $\e_n.$
We use capitals for matrices, e.g. adjacency matrix $G$ of network $\Grid.$
When the considered vectors and matrices grow exponentially in number of people $N,$ and hence suffer from the curse of dimensionality, we highlight it using bold font, e.g. $\p$ for high--dimensional probability distribution function, and $\A$ for the matrix of CME.

\section{Methods}  \label{sec:methods}     
 \subsection{Forward problem: ${\varepsilon}$--SIS epidemic on network} \label{sec:eSIS}        
 We consider the $\eps$--SIS model of the contact process, which is a variation of a classical susceptible--infected--susceptible (SIS) model, allowing for every node to self--infect with rate $\eps.$
 This process was originally proposed by Hill et al. to describe the spread of emotions in social network~\cite{hill-sis-2010}.
 Mieghem et al. studied analytical properties of the model for fully connected networks~\cite{mieghem-sis-2012,mieghem-sis-2022}.
 Zhang et al. extended this study to arbitrary networks and found conditions under which the equilibrium distribution can be accurately approximated by a scaled SIS process, gaining useful insights on vulnerability of the population~\cite{zhang-sis-2017}.
 
 The classical SIS model has an absorbing state where all nodes are susceptible (i.e. the network is fully healthy), but due to spontaneous self--infections the $\eps$--SIS model does not have an absorbing state, hence the epidemics lasts forever.
 This property allows us to observe the epidemics for sufficiently long time and eventually collect the dataset which is large enough to ensure the required contrast for the Bayesian optimisation.

 We consider a $\eps$--SIS epidemic on a unweighted simple network 
 \(
 \Grid = (\Nodes,\Edges),
 \)
 which is a set of nodes (representing people, or agents) 
 \(
 \Nodes = \{1,2,\ldots,N\}
 \)
 and links (or edges, representing contacts between agents)
 \(
 \Edges = \{(m,n):\: m\in\Nodes, n\in\Nodes, \: m\neq n\}.
 \)
 We additionally assume that the contacts are bidirectional, i.e. 
 \(
  (m,n) \in\Edges \:\Leftrightarrow\: (n,m)\in\Edges,
 \)
 which allows us to introduce a symmetric adjacency relation
 \(
 m\sim n \:\Leftrightarrow\: (m,n)\in\Edges
 \)
 for the connections.
 
\begin{figure}[t]
 \begin{center}
   \includegraphics[width=.95\textwidth]{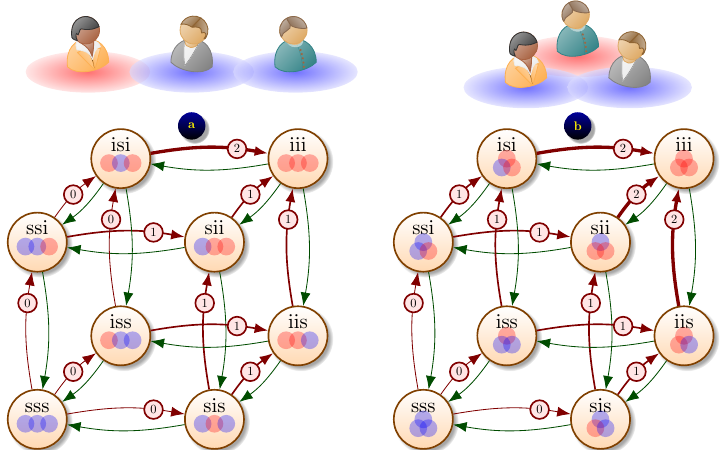}
  \end{center}
 \caption{Markov transitions between network states:
            (a) ${\varepsilon}$--SIS epidemic on a chain of $N=3$ people;
            (b) ${\varepsilon}$--SIS epidemic in a fully connected network of $N=3$ people.
         On the graph, 
           green arrows denote recovery process with rate $\gamma,$ and
           red arrows with a circled number $k$ denote infection process with rate $k\beta+\varepsilon.$
            }
 \label{fig:CME}
\end{figure}

 Each node can be in one of two states,
 \(
 x_n \in \X 
  = \{\text{susceptible}, \text{infected}\}
  = \{0,1\},
  \)
  for $n\in\Nodes.$
The state of the whole network is therefore a vector 
\(
 \x = \begin{pmatrix}x_1 & x_2 & \ldots & x_N \end{pmatrix}^T \in \X^N.
\)
We consider the system dynamics as a continuous--time Markov jump process on the state space $\Omega=\X^N.$ 
The following two types of transitions (or reactions), infection and recovery, occur independently and at random according to the Poisson process with the following rates
\begin{equation}\label{eq:reactions}
 p_{\x\to\y} =
  \begin{cases}
   p_{\x\to\y}^\inf, &\text{if $\exists n\in\Nodes:\, \y=\x+\e_n$;} \\  
   p_{\x\to\y}^\rec, &\text{if $\exists n\in\Nodes:\, \y=\x-\e_n$;} \\
    0,           &\text{otherwise,}
  \end{cases}
\end{equation}
where $\e_n\in\Real^N$ is the $n$--th unit vector.
For simplicity, we assume that the recovery rates $p_{\x\to\y}^\rec=\gamma$ are the same for all nodes of the network.
In the classical SIS model, the infection rate for the susceptible node $x_n=0$ is proportional to the number of its infected neighbours,
 \(
 I_n(\x) = \left| \{ m\in\Nodes: m\sim n, x_m=1 \} \right|,
 \) 
and the per--contact rate $\beta,$ which we also consider the same across all network.
In the $\eps$--SIS model, an additional infection rate $\eps$ is introduced to describe possible infection through contacts with the external, off-the-network, environment.
Hence, the infection rate is
 \(
 p_{\x\to\y}^\inf=I_n(\x)\beta+\eps.
 \)
Examples of the Markov transition graph are shown in Fig.~\ref{fig:CME}.

The stochastic properties of the system are described as probabilities of network states
\(
p(\x,t) = \Prob(\text{system is in state $x$ at time $t$}).
\)
The system dynamics is written as a system of ordinary differential equations (ODEs),
known as \emph{Markovian master equation}~\cite{vankampen-stochastic-1981,chen-stoch-sir-2005}, Chapman or \emph{forward Kolmogorov equations}:
\begin{equation}\label{eq:cme}
 p'(\x,t) 
  = \sum_{\y\in\X^N} 
        \left( 
          p_{\y\to\x} \cdot p(\y,t) - p_{\x\to\y}\cdot p(\x,t)
        \right), \qquad \x\in\X^N,
\end{equation}
subject to initial conditions  
\(
p(\x_0,0) = 1
\) 
for the initial state $\x=\x_0$ and 
\(
p(\x,0) = 0
\)
otherwise.
The number of ODEs scales as $2^N,$ making traditional numerical solvers struggle for even moderate values of $N.$

\subsection{Inverse problem: Bayesian inference of the network} \label{sec:Bayes}   
Our goal is to infer the most probable contact network 
\(
\Grid=(\Nodes,\Edges)
\)
from the observed data
\(
\Data = \{t_k,\x(t_k)\}_{k=0}^K.
\)
According to the Bayes theorem~\cite{box-bayes-1973}, 
 \( 
 \Prob(\Grid | \Data) = \frac{\Prob(\Data | \Grid) \Prob(\Grid)}{\Prob(\Data)},
 \)
where 
  $\Prob(\Grid)$ is the \emph{prior} probability distribution for the grid,
  $\Prob(\Data|\Grid)$ is the \emph{likelihood} of the observed data as a function of the grid $\Grid,$
  $\Prob(\Data)$ is the probability of the observed data, and
  $\Prob(\Grid|\Data)$ is the posterior probability of the grid given the data.

\begin{figure}[t]
 \begin{center}
   \includegraphics[width=.95\textwidth]{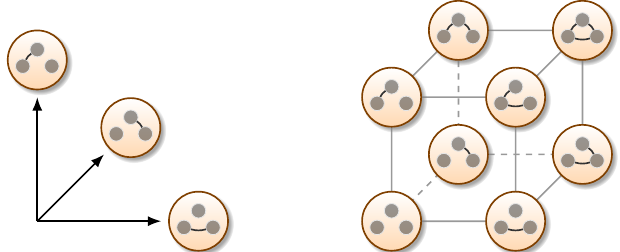}
 \end{center}
 \caption{The set of all possible networks with $N=3$ nodes is a binary hypercube in dimension $\tfrac12 N(N-1)=3.$}
 \label{fig:grids}
\end{figure}

 The connectivity network $\Grid$ can be described by its adjacency matrix
 \(
 G = \left[g_{m,n}\right]_{m,n\in\Nodes}
 \)
 with 
 \(
 g_{m,n} = 1 \:\Leftrightarrow\: (m,n)\in\Edges
 \)
 and
 \(
 g_{m,n} = 0,
 \)
 otherwise.
 Since $\Grid$ is simple, $G$ is a binary symmetric matrix, $G=G^T,$ with zero diagonal, $\diag(G)=0.$
 It is easy to see that a grid $\Grid$ can be described by $\tfrac12 N(N-1)$ independent binary variables $g_{m,n}\in\Binary=\{0,1\}$ with $m,n\in\Nodes$ and $m>n,$ representing the states (on/off) of possible edges $(m,n)\in\Edges.$
 Therefore, for a fixed number of nodes $|\Nodes|=N,$ the set of all possible networks 
 \(
 \Grids = \{\Grid: ~g_{m,n}=g_{n,m}\in\Binary, ~m,n\in\Nodes, ~m>n\}
 \)
 has cardinality 
 \(
 |\Grids| = 2^{N(N-1)/2}.
 \)
Note that $\Grids$ is isomorphic to $\Binary^{N(N-1)/2}$, the set of adjacency elements $g_{m,n}$.
The structure of this set is illustrated in Fig.~\ref{fig:grids}.

 Assuming that nodes are known and no prior information of the edges $\Edges$ is available, we take the uniform prior distribution,
 \(
 \Prob(\Grid) = 2^{-N(N-1)/2}.
 \)
Although $\Prob(\Data)$ is unknown, it does not affect the optimisation problem, as
\(
\max_{\Grid} \Prob(\Grid|\Data) = \frac{2^{- N(N-1)/2}}{\Prob(\Data)} \max_{\Grid} \Prob(\Data|\Grid), 
\)
hence
\(
\arg\max_{\Grid} \Prob(\Grid|\Data) = \arg\max_{\Grid} \Prob(\Data|\Grid).
\)

Due to the Markovian property of the system dynamics, the likelihood can be expanded as follows,
\begin{equation}\label{eq:markov}
 \begin{split}
  \Like(\Grid) 
          & = \Prob(\Data | \Grid) 
            = \Prob( X(t_1)=\x_1, \cdots, X(t_K)=\x_K | \Grid)
     \\   & = \Prob( X(t_1)=\x_1 | X(t_0)=\x_0, \Grid) \cdots \Prob( X(t_K)=\x_K | X(t_{K-1})=\x_{K-1}, \Grid), 
     \\   & = \prod_{k=1}^K \underbrace{\Prob( X(t_k)=\x_k | X(t_{k-1})=\x_{k-1}, \Grid)}_{\Prob(\x_{k-1}\to \x_k | \Grid)},
 \end{split}
\end{equation}
where $X(t)\in\X^N$ are random variables describing the network states during its stochastic evolution, and
 the time sequence $\{t_k\}_{k=0}^K$ is monotonically increasing.
We see that the likelihood is a product of transition probabilities 
\(
\Prob(\x_{k-1}\to \x_k | \Grid) = \Prob( X(t_k)=\x_k | X(t_{k-1})=\x_{k-1}, \Grid),
\)
which are the probabilities for the system to evolve from the state $\x_{k-1}$ to $\x_k$ over the time period $[t_{k-1},t_k].$
The (black--box) Bayesian network inference therefore boils down to likelihood optimisation,
\begin{equation}\label{eq:loglike}
 \begin{split}
  \Grid_{\text{opt}} & = \arg\max_{\Grid\in\Grids} \log\Like(\Grid)
           = \arg\max_{\Grid\in\Grids} \sum_{k=1}^K  \log\Prob( \x_{k-1} \to \x_k | \Grid).
 \end{split}
\end{equation}

To compute a single log--likelihood $\log\Like(\Grid)$ in~\eqref{eq:loglike}, we need to solve $K$ forward problems, i.e.
  estimate the chance of arriving to the state $\x_k$ from the state $\x_{k-1}$ over the period of time $t\in[t_{k-1},t_k]$
  for $k=1,\ldots,K.$
To find the optimal network $\Grid_{\text{opt}}$, we may need to compute a large amount of log--likelihoods for different networks $\Grid,$
  hence the efficiency of the forward solver is crucial to make the optimisation procedure feasible.
This rules out a possibility of solving~\eqref{eq:cme} directly due to the curse of dimensionality.

\subsection{Stochastic simulation algorithms for forward problem} \label{sec:SSA}         
Traditionally, probabilities $p(\x,t)$ are estimated using the Stochastic Simulation Algorithm (SSA)~\cite{gillespie-ssa-1976}, or some more efficient (e.g. multilevel) Monte Carlo simulations of the realisations of the model~\cite{Gillespie2001,hemberg-perfect-sampling-2007,AndersonHigham-MLCME-2012,Yates-MLCME-2016}.
Essentially, these methods sample $\Nssa$ random walks through the state space $\Omega=\X^N$ starting at the initial state $\x_{k-1},$  count the number $\nssa$ of trajectories that end up in the state $\x_k$ by the time $t_k,$ and estimate the target probability as a frequency,
\begin{equation}\label{eq:ssa}
 \Prob( \x_{k-1} \to \x_k | \Grid) 
   \approx \tilde\Prob( \x_{k-1} \to \x_k | \Grid) 
   = \frac{\nssa}{\Nssa}.
\end{equation}
The cost of sampling each trajectory does not grow exponentially with $N$ which makes these methods free from the curse of dimensionality.
However, the convergence of these algorithms is not particularly fast, with typical estimates
\[
 \mathrm{err} 
  = \left| 
    \Prob( \x_{k-1} \to \x_k | \Grid) 
     -
     \tilde\Prob( \x_{k-1} \to \x_k | \Grid) 
   \right|
   \leq
   c \Nssa^{-\delta},
\]
with $0.5\leq\delta\leq 1$ depending on a particular method.
This is usually sufficient to estimate large probabilities and main statistics (such as mean and variance) of the process.
However, if the probability
\(
p = \Prob( \x_{k-1} \to \x_k | \Grid) 
\)
is small,
to ensure the desired relative precision
\(
| p - \tilde p | \leq \epsilon p,
\)
the number of samples should satisfy
\(
c \Nssa^{-\delta} \leq \epsilon p,
\)
or 
\(
\Nssa \geq (\epsilon p  / c)^{-1/\delta}.
\)
In practice this means that estimation of rare events with probabilities $p\lesssim 10^{-6}$ with these algorithms can be prohibitively expensive.

If the computational budget of $\Nssa$ trajectories is exhausted and none of them arrived at $\x_k$, then $\nssa=0$ and $\tilde p=0,$ i.e. the event is not resolved with the algorithm and is considered impossible.
If this happens for at least one transition from state $\x_{k-1}$ to $\x_k$ for some $k=1,\ldots,K,$ then the whole likelihood $\Like(\Grid)=0$ for the given network $\Grid.$
In practical computations this issue can occur for most networks $\Grid\neq\Grid_\star$ except the `ground truth' network and its close neighbours.
The limited computation budget for the forward problem therefore leads to flattening of the high--dimensional landscape of the likelihoods, wiping off the structural information that should be used to navigate the optimisation algorithm towards the solution of~\eqref{eq:loglike}.
This motivates the development of more accurate methods for the forward problem, such as the tensor product approach that we discuss in the following subsection.

\subsection{Tensor product algorithms for forward problem} \label{sec:TT}           
To find the probabilities composing the likelihood~\eqref{eq:loglike}, we can solve the system of ODEs~\eqref{eq:cme} for the probabilities of the network states.
This system is commonly known as the chemical master equation (CME) and consists of $|\X^N|=2^N$ equations and unknowns, hence traditional solvers suffer from the curse of dimensionality.
To mitigate this problem, different approaches were used, including
  sparse grids~\cite{hegland-cme-2007},
  adaptive finite state projections~\cite{munsky-fsp-2006,jahnke-wavelet-cme-2010,Cao-FSP-2016},
  radial basis functions~\cite{Schuette-RBF-CME-2015},
  neural networks~\cite{Khammash-NN-CME-2021,Grima-CME-NN-2022},
  and tensor product approximations, such as 
       canonical polyadic (CP) format~\cite{jahnke-cme-2008,Ammar-cme-2011,hegland-cme-2011},
       and more recently tensor train (TT) format~\cite{kkns-cme-2014,dkh-cme-2014,ds-amen-2014,d-tamen-2018,Sidje-TT-CME-2017,Dinh-QTT-CME-2020,Ion-TT-CME-2021,Schuette-CME-CO-2016,ds-ttsir-2024}.
Here we briefly describe the tensor train approach used in our recent paper~\cite{ds-ttsir-2024}.

\begin{figure}[t]
 \begin{center}
   \includegraphics[width=.95\textwidth]{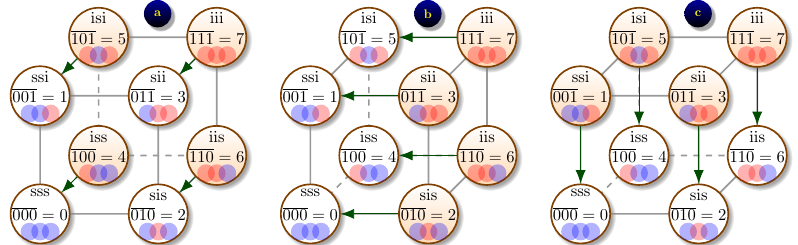}
\end{center}
 \caption{The tensor product structure of recovery transitions \( [ p^\rec_{\x\to(\x-\e_n)} ]_{\x\in\X^N}  \) is illustrated for population of $N=3$ people. 
 Recovery takes place on individual nodes and hence does not depend on contact network.
 In each panel, highlighted states $\x$ are where \( p^\rec_{\x\to(\x-\e_n)} = \gamma \), indicating that person $n$ is infected and can recover; this is also shown by green arrows. 
 Non--highlighted states correspond to \( p^\rec_{\x\to(\x-\e_n)} = 0 \).
         (a) $n=1,$
         (b) $n=2,$
         (c) $n=3.$
        }
  \label{fig:tensor}
\end{figure}

First, we note that among $2^N$ reaction rates $p_{\y\to\x}$ in~\eqref{eq:cme} only $2N$ are non--zero, according to~\eqref{eq:reactions}:
\begin{equation}\label{eq:cme2}
 \begin{split}
 p'(\x,t) 
  & = \sum_{n=1}^N
        \left( 
          p^\inf_{(\x-\e_n)\to\x} p(\x-\e_n,t) + p^\rec_{(\x+\e_n)\to\x} p(\x+\e_n,t)  
      \right. \\ & \left.
      \qquad\qquad\qquad
          - \left(p^\inf_{\x\to(\x+\e_n)}+ p^\rec_{\x\to(\x-\e_n)}\right)p(\x,t)
        \right),
 \end{split}
\end{equation}
for \( \x\in\X^N. \)
Let us now uncover the tensor product structure of the matrix of this CME. 
For this we assume that $2^N$ network states $\x\in\X^N$ are ordered lexicographically, i.e. the state $\x=(x_1,\cdots,x_N)^T$ has index
\[
 \overline{\x}
 = \overline{x_1x_2\ldots x_N}
 = 2^{N-1} x_1
 + 2^{N-2} x_2
 + \cdots
 + 2^0 x_N,
\]
which corresponds to how binary numbers are written in big--endian notation, e.g.
   \(\overline{000} = 0,\)
   \(\overline{001} = 1,\)
   \(\overline{010} = 2,\)
   \(\overline{100} = 4,\)
   \(\overline{101} = 5,\)
   \(\overline{111} = 7,\) 
   see also Fig.~\ref{fig:tensor}.
When the probability distribution function (p.d.f.) vector
 \(
 \p(t)=\begin{bmatrix}p(\x,t)\end{bmatrix}_{\x\in\X^N}
 \)
is composed, we place the probability $p(\x,t)$ in position $\overline{\x}$,  assuming 0-based indexing scheme, i.e. vector indices enumerated from $0$ up to $2^N-1.$
If 1-based indexing scheme is used, we place $p(\x,t)$ in position $\overline{\x}+1.$

Using \emph{indicator} function 
\[
 \Indicator{\text{\emph{condition}}}
  = \begin{cases}
   1, & \text{if \emph{condition} is true} \\
   0, & \text{if \emph{condition} is false},
  \end{cases}
\]
we can write
\(
 p^\rec_{\x\to(\x-\e_n)} = \gamma \Indicator{x_n=1},
\)
and
\(
p^\inf_{\x\to(\x+\e_n)} = (\eps + \beta \sum_{m\sim n}\Indicator{x_m=1})\Indicator{x_n=0}.
\)
Collecting these reaction rates in vectors of size $2^N$, and using the big--endian lexicographic ordering as explained above, we obtain tensor product decomposition
\begin{equation}\label{eq:ttrec}
 \begin{split}
 \begin{bmatrix} p^\rec_{\x\to(\x-\e_n)} \end{bmatrix}_{\x\in\X^N}
    & = \gamma 
         \evec \otimes \cdots \otimes \evec 
         \otimes \ivec \otimes 
          \evec  \otimes \cdots \otimes \evec, \\
 \end{split}
\end{equation}
where 
   \( \ivec=\begin{pmatrix}0&1\end{pmatrix}^T \) appears in position $n,$
   \( \evec=\begin{pmatrix}1&1\end{pmatrix}^T \) appear elsewhere.
 The tensor product structure is illustrated for $N=3$ in Fig.~\ref{fig:tensor}.
 For example, the full vector of recovery transitions corresponding to recovery of person $n=1$ is
 \begin{equation}\label{eq:example}
  \begin{split}
    \begin{bmatrix} p^\rec_{\x\to(\x-\e_1)} \end{bmatrix}_{\x\in\X^N} 
          & = \gamma \ivec \otimes \evec \otimes \evec
      \\  & = \gamma  \begin{pmatrix}0&1\end{pmatrix}^T \otimes \begin{pmatrix}1&1\end{pmatrix}^T  \otimes \begin{pmatrix}1&1\end{pmatrix}^T 
       \\  & = \gamma  \begin{pmatrix}0 & 0 & 0 & 0 & 1 & 1 & 1 & 1\end{pmatrix}^T.
  \end{split}
 \end{equation}
 We can see that this transition is possible from states \( \x=\begin{pmatrix}x_1 & x_2 & x_3\end{pmatrix} \) with $x_1=1,$ which are located in positions
  \( \overline{100} = 4, \)
  \( \overline{101} = 5, \)
  \( \overline{110} = 6, \)
  \( \overline{111} = 7 \)
  of the vector (assuming 0-based indexing), in agreement to eq.~\eqref{eq:example} and Fig.~\ref{fig:tensor}(a).

Similarly,
\begin{equation}\label{eq:ttinf}
 \begin{split}
 \begin{bmatrix} p^\inf_{\x\to(\x+\e_n)} \end{bmatrix}_{\x\in\X^N}
    & = \eps 
         \evec \otimes \cdots \otimes \evec 
         \otimes \svec \otimes 
          \evec  \otimes \cdots \otimes \evec 
   \\&   + \beta \sum_{m\sim n} 
         \evec \otimes \cdots \otimes \evec 
         \otimes \svec \otimes 
          \evec  \otimes \cdots \otimes \evec 
         \otimes \ivec \otimes 
         \evec  \otimes \cdots \otimes \evec,
 \end{split}
\end{equation}
where 
   $\svec=\begin{pmatrix}1&0\end{pmatrix}^T$ appears in position $n,$ 
   $\ivec=\begin{pmatrix}0&1\end{pmatrix}^T$ appear in positions $m\sim n,$ 
   $\evec=\begin{pmatrix}1&1\end{pmatrix}^T$ appear elsewhere.
To complete the expansion for the right--hand side of~\eqref{eq:cme2}, we need to express the shifted state 
\(
p(\x-\e_n,t)
\)
as a sum over probabilities $p(\y,t)$ as follows
\begin{equation}\label{eq:shift}
 \begin{split}
   p(\x-\e_n,t) 
  & = \sum_{\y\in\X^N}
             \Indicator{x_1=y_1} \cdots
             \Indicator{x_{n-1}=y_{n-1}} \cdot 
             \Indicator{x_n-1=y_n} \cdot 
             \Indicator{x_{n+1}=y_{n+1}} \cdots
             \Indicator{x_N=y_N} 
             \cdot p(\y,t),
             \\
 \begin{bmatrix}  p(\x-\e_n,t) \end{bmatrix}_{\x\in\X^N}
    & =  
             \underbrace{\left(
             \Id \otimes \cdots \otimes \Id \otimes J^T \otimes \Id \otimes \cdots \otimes \Id
             \right)}_{\J_n^T}
             \p(t),
 \end{split}
\end{equation}
where the \emph{shift matrix}
  \(
  J^T=\left(\begin{smallmatrix} \zero& \zero \\ 1 &\zero \end{smallmatrix}\right)
  \)
  appears in position $n,$ 
  and identity matrices
  \(
  \Id=\left(\begin{smallmatrix} 1 & \zero \\ \zero & 1 \end{smallmatrix}\right)
  \) 
  appear elsewhere.
  Similarly, 
\( 
 \begin{bmatrix}  p(\x+\e_n,t) \end{bmatrix}_{\x\in\X^N} = \J_n \p
\)
with
\(
 \J_n = \Id \otimes \cdots \otimes \Id \otimes J \otimes \Id \otimes \cdots \otimes \Id.
\)
Combining the above, we collect all equations of~\eqref{eq:cme2} in a vectorised CME
\begin{equation}\label{eq:CME}
 \p'(t) = \A \p(t), \qquad \p(0)=\p_0,
\end{equation}
where the $2^N\times 2^N$ matrix  $\A$ admits the following tensor product form:
\begin{equation}\label{eq:Acp}
 \begin{split}
  \A & =
  \gamma\sum_{n=1}^N 
         \Id \otimes \cdots \otimes \Id
         \otimes  (J-\Id) \hat I \otimes 
          \Id  \otimes \cdots \otimes \Id 
   \\ & +  
  \eps\sum_{n=1}^N 
         \Id \otimes \cdots \otimes \Id
         \otimes  (J^T-\Id) \hat S \otimes 
          \Id  \otimes \cdots \otimes \Id 
   \\ & +  
  \beta\sum_{n=1}^N \sum_{m\sim n}
         \Id \otimes \cdots \otimes \Id
         \otimes  (J^T-\Id) \hat S \otimes 
          \Id  \otimes \cdots \otimes \Id 
         \otimes \hat I \otimes 
         \Id  \otimes \cdots \otimes \Id,
 \end{split}
\end{equation}
where 
\(
\hat I = \diag(\ivec) = 
      \left(\begin{smallmatrix} \zero & \zero \\ \zero & 1 \end{smallmatrix}\right)
\)
and
\(
\hat S = \diag(\svec) = 
      \left(\begin{smallmatrix} 1 & \zero \\ \zero & \zero \end{smallmatrix}\right).
\)
This so-called canonical polyadic (CP)~\cite{hitchcock-sum-1927,harshman-parafac-1970,cc-parafac-1970} form represents the CME matrix $\A$ as a sum of $(2N+|\Edges|)$ elementary tensors, each of which is a direct product of $N$ small $2\times 2$ matrices, acting on a single node of the network only.
Hence, the storage for $A$ reduces from $\O(2^N \langle k \rangle)$ down to $\O((2N+\langle k \rangle)N)$ elements, where $\langle k \rangle = |\Edges|/|\Nodes|$ denotes the average degree of $\Grid.$
The curse of dimensionality for the matrix is therefore removed.

To remove the exponential complexity in solving~\eqref{eq:CME}, we need to achieve a similar compression for the p.d.f.~$\p(t)=[p(\x,t)]_{\x\in\X^N},$ for which we employ the \emph{tensor train} (TT) format~\cite{osel-tt-2011}.
 \begin{equation}\label{eq:tt}
  \p \approx \tilde\p 
      = \sum_{\alpha_0,\ldots,\alpha_N=1}^{r_0,\ldots,r_N} 
             \pcore^{(1)}_{\alpha_0,\alpha_1} 
             \otimes \cdots \otimes 
             \pcore^{(n)}_{\alpha_{n-1},\alpha_n} 
             \otimes \cdots \otimes 
             \pcore^{(N)}_{\alpha_{N-1},\alpha_N}.
 \end{equation}
Here, the $r_{n-1} \times 2 \times r_n$ factors
  \(
   \pcore^{(n)} = \left[ \pcore^{(n)}_{\alpha_{n-1},\alpha_n}(x_n) \right],
  \)
   $n=1,\ldots,N,$ 
   are called \emph{TT cores}, and
 the ranges of the summation indices $r_0,\ldots,r_N$ are called \emph{TT ranks}.
Each core $\pcore^{(n)}$ contains information related to node $n$ in the network, 
   and the summation indices $\alpha_{n-1},\alpha_n$ of core $\p^{(n)}$ link it to cores $\pcore^{(n-1)}$ and $\pcore^{(n+1)}.$
The matrix--vector multiplication can be performed fully in tensor product format.
Using recently proposed algorithms~\cite{ds-amen-2014,d-tamen-2018} the linear system of ODEs~\eqref{eq:CME} can be solved fully in the TT format avoiding the curse of dimensionality, as explained in details in~\cite{ds-ttsir-2024}.

\subsection{Ordering of network nodes for faster forward problem solving} \label{sec:sort}  
 The TT decomposition of probability functions exhibits low ranks when distant (with respect to their position in the state vector) variables are weakly correlated; see e.g. \cite{rdgs-tt-gauss-2022} for a rigorous analysis of this for the multivariate normal probability density function.
Numerical approaches to order the variables in such a way include 
  a greedy complexity optimisation over a reduced space of permutations~\cite{ballani-adapttree-2014,nouy-tree-learning-2022}, 
  using gradients to compute a Fisher--type information matrix and its eigenvalue decomposition to sort or rotate the variables~\cite{cdo-cond-dirt-2023}, and
  sorting the variables according to the Fiedler vector of the network~\cite{legeza-graph-opt-2011}.
Since in our case the variables are discrete, we adopt the latter approach.

We consider the Laplacian matrix of the network $\Grid,$ defined as follows
\begin{equation}\label{eq:laplace}
L = \diag( G \e ) - G,
\end{equation}
where 
\(
G\in\Binary^{N\times N}
\)
is the adjacency matrix of $\Grid,$ and 
\(
\e = \begin{pmatrix} 1 & 1 & \cdots & 1 \end{pmatrix}^T \in \Binary^N.
\)
We are particularly interested in the \emph{Laplacian spectrum} of $\Grid,$ i.e. solutions to the eigenvalue problem
\(
 L \u = \lambda \u.
\)
It is easy to see that since $G=G^T$ we also have $L=L^T,$ hence the spectrum is real, $\lambda\in\Real.$
We also note that 
\(
L=\left[\ell_{m,n}\right]_{m,n\in\Nodes}
\) 
is diagonally dominant, since
\(
 \ell_{m,m}= \sum_{n\in\Nodes} g_{m,n} \geq 0,
\)
and
\(
\ell_{m,n} = -g_{m,n} \leq 0
\)
for $m\neq n,$
hence
\(
|\ell_{m,m}| = \sum_{m\neq n} |\ell_{m,n}|.
\)
Since $\diag(L)\geq0$ and $L$ is symmetric and diagonally dominant, the Laplacian is positive semi-definite, hence its eigenvalues are nonnegative,
\(
\lambda_{n-1}  
 \geq \lambda_{n-2} 
 \geq \ldots
 \geq \lambda_1 
 \geq \lambda_0
 \geq 0.
\)
It is easy to see that $L\e = 0,$ hence $\lambda_0=0$ is the lowest eigenvalue with the corresponding eigenvector $\u_0=\e.$

The second eigenvalue, which we denote $\lambda_1,$ is called the \emph{algebraic connectivity} of $\Grid$ or \emph{Fiedler value}.
It was known since~\cite{fiedler-value-1973} that $\lambda_1=0$ if and only if $\Grid$ is disconnected.
This is a particularly easy scenario for epidemiological modelling.
If $\Grid$ consists of two disjoint networks, $\Grid_1$ and $\Grid_2,$ then nodes from $\Grid_1$ and $\Grid_2$ can not affect each other.
The random variables associated with those nodes are therefore independent, i.e. if $X_1\in\Grid_1$ and $X_2\in\Grid_2$ then 
\(
\Prob(X_1, X_2) = \Prob(X_1) \Prob(X_2).
\)
This means that if we order the variables $x_1,\ldots,x_N$ in such a way that the first block $x_1,\ldots,x_m\in\Grid_1$  and the second block $x_{m+1},\ldots,x_N \in \Grid_2$ do not overlap,
 then the TT format~\eqref{eq:tt} for the p.d.f. $p(x_1,\ldots,x_N)$
 will have the TT rank $r_m=1$ for the connection separating $\Grid_1$ and $\Grid_2$.

These geometric properties of the network can be estimated from the eigenvector $\v = \u_1,$ also known as \emph{Fiedler vector}.
In the pioneering paper~\cite{fiedler-vector-1975} it was related to finding an optimal cut in the network $\Grid.$
It was generalised to reducing the envelope (or the bandwidth) of the adjacency matrix $G$ in~\cite{barnard-fiedler-1993},
  and later to finding orderings of variables for which TT decomposition~\eqref{eq:tt}, and related MPS and DMRG representations in quantum physics~\cite{schollwock-2011}, have lower TT ranks~\cite{legeza-graph-opt-2011}.
The Fiedler vector can be computed by minimising the \emph{Rayleigh quotient}
\(
\v = \arg\min_{\v\perp\e} \v^T L \v / \|v\|^2,
\)
also known as the \emph{Courant minimax} principle, orthogonally to $\u_0=\e.$
Following~\cite{barnard-fiedler-1993}, we use the Fiedler vector to define a one--dimensional embedding of the graph to a linear chain.
Let 
\(
\sigma\in\Permutations_n
\)
be the permutation vector of the set of nodes $\Nodes$ such that $(\v_\sigma)$ is ordered, i.e.
\(
\v_{\sigma_1}\geq\v_{\sigma_2}\geq\cdots\geq\v_{\sigma_N},
\)
or equivalently in the ascending order.
Following~\cite{legeza-graph-opt-2011}, the same permutation of variables also reduces the TT ranks of the TT decomposition~\eqref{eq:tt}.
In particular, it groups together variables corresponding to independent subnetworks.
Hence, we compute this permutation and adopt its order of variables before solving the forward problem~\eqref{eq:CME} using tensor product algorithms.

\subsection{Algorithms for Bayesian inverse problem} \label{sec:MCMC}  
Since the full grid search is unfeasible for even moderate networks,
we adopt the Metropolis--Hastings Markov Chain Monte Carlo (MCMC) method to approach the maximum of $\Like(\Grid)$.
The method is depicted in Algorithm~\ref{alg:mcmc}.
We need to choose a \emph{proposal distribution} $q(\hat \Grid | \Grid)$ which is tractable for sampling a new state of the network $\hat \Grid$, given the current network $\Grid$.
In each iteration, given the current network $\Grid_i$, we sample a new proposal $\hat \Grid$ and accept or reject it with probability based on the Metropolis--Hastings ratio, forming a Markov Chain of network configurations $\Grid_0,\Grid_1,\ldots$
After the Markov Chain is computed, we return the sample of this chain with the maximal likelihood.
\begin{algorithm}[htb]
\centering
\caption{MCMC algorithm for the likelihood maximisation}
\label{alg:mcmc}
\begin{algorithmic}[1]
 \State Choose the proposal distribution $q(\hat \Grid | \Grid)$, initial network $\Grid_0$ and length of the chain $\neval$.
 \For{$i=0,\ldots,\neval-1$}
   \State Sample a new proposal network $\hat \Grid \sim q(\hat \Grid | \Grid_{i})$.
   \State Compute the Metropolis--Hastings ratio $h(\hat \Grid, \Grid_i) = \frac{\Like(\hat \Grid)}{\Like(\Grid_i)} \frac{q(\Grid_i | \hat \Grid)}{q(\hat \Grid | \Grid_{i})}$. \Comment{As in \eqref{eq:mh}.}
   \State Sample a uniformly distributed random number $r \sim \Uniform(0,1)$.
   \If{$r<\min(h(\hat \Grid, \Grid_i), 1)$}
     \State Accept the proposal by setting $\Grid_{i+1} = \hat \Grid$.
   \Else
     \State Reject the proposal by setting $\Grid_{i+1} = \Grid_i$.
   \EndIf
 \EndFor \\
 \Return $\Grid_\star = \arg\max_{i=0,\ldots,\neval} \log\Like(\Grid_i)$.
\end{algorithmic}
\end{algorithm}
This algorithm is known to converge to the true distribution $\frac{1}{Z}\Like(\Grid)$ (where $Z$ is the normalising constant) under mild assumptions~\cite{RobertsRosenthal2011}.
We implement two proposal distributions.
\begin{enumerate}
 \item Choose one link in $\Grid_i$ uniformly at random and toggle its state (on/off). Since there are $N(N-1)/2$ possible links to toggle, $q(\hat \Grid | \Grid) = \frac{2}{N(N-1)}$ independently of $\hat \Grid$ and $\Grid$, and hence cancels in the Metropolis-Hastings ratio. We will call Algorithm~\ref{alg:mcmc} with this proposal ``MCMC-R'', since it samples the links with Replacement.
 \item Every $N(N-1)/2$ iterations (i.e. when $\mod(n,N(N-1)/2)=0$), sample a random permutation vector 
 \(
  \sigma\in\Permutations_{N(N-1)/2}
 \)
  of the set $\{1,\ldots,N(N-1)/2\},$ and in the next $N(N-1)/2$ iterations toggle links in the order prescribed by $\sigma$.
 This is still a valid MCMC algorithm with a constant proposal distribution, but now with respect to $\sigma$, corresponding to a collection of networks in consecutive update steps, $(\Grid_{i+1}, \ldots, \Grid_{i+N(N-1)/2})$.
 In our numerical experiments reported in Section~\ref{sec:results}
 this algorithm sometimes increased the likelihood faster in terms of the individual link changes (and hence computing time), and
 resulted in a better grid reconstruction.
 In each block of $N(N-1)/2$ iterations, this algorithm proposes link changes without replacement \cite{Shah-noR-2018}.
 For this reason, we will call this algorithm ``MCMC-noR'' (for ``no Replacement'').
\end{enumerate}

Cancelling the constant proposal probability, and using log-likelihoods to avoid numerical over- and under-flow errors, we
can rewrite the Metropolis--Hastings ratio as
\(
 h(\hat \Grid, \Grid_i) = e^{\ln\Like(\hat\Grid) - \ln\Like(\Grid_i)}.
\)
 We can further modify this formula to gain more control on performance of MCMC.
 Specifically, we introduce the \emph{temperature}, or \emph{tempering} parameter $\Temp$ as follows:
\begin{equation}\label{eq:mh}
 h(\hat \Grid, \Grid_i) = \exp\left(-\frac{\ln\Like(\Grid_i) - \ln\Like(\hat\Grid)}{\Temp}\right).
\end{equation}
We note that this formula resembles the Boltzmann distribution, also know as Gibbs distribution, which is used in statistical and theoretical physics and describes probability to observe particles in the ground and exited energy states.
Similarly, the parameter $\Temp$ controls how often MCMC accepts a network with worse likelihood that the current one, which in order affects the convergence of the algorithm and its ability to get out of local optima.
If a new grid $\hat\Grid$ has better likelihood than the current grid $\Grid_i,$ it is always accepted.
Otherwise, it is accepted only with probability
 \(
 p=\Like(\hat\Grid)/\Like(\Grid_i)<1
 \)
 for $\Temp=1$.
 For $\Temp>1,$ this probability becomes $p^{1/\Temp} > p,$ which increases the probability that the new grid is accepted, encouraging MCMC to escape a local maximum.

\subsection{Choosing initial guess for optimisation} \label{sec:init}  
A good initial guess $\Grid_0$ for the contact network can significantly improve the computational efficiency by reducing the number of steps required by the optimisation algorithm to converge towards the optimum $\Grid_{\text{opt}},$ but also by simplifying the forward problems and hence reducing the computational time required to evaluate each likelihood~$\Like(\Grid)$ in~\eqref{eq:loglike}.
Here we present a simple algorithm to generate an initial guess using the given nodal time series data
\(
\Data = \{t_k,\x_k\}_{k=0}^K.
\)
By comparing each next state $\x_k$ against a previous one $\x_{k-1}$ for $k=1,\ldots,K$ node-by-node, we observe events of two possible types:  
  infected nodes that become susceptible (recovery), and
  susceptible nodes becoming infected (infection).
Recoveries are single--node events and provide no information on the network connectivity.
In contrast, infections are two--node events that occur when a susceptible node is connected to an infected node.
Therefore, any node $m$ that was or became infected during $t\in[t_{k-1},t_k]$ could have infected any connected susceptible node $n$ that became infected during the same time interval.

Thus, we compute the connectivity \emph{scores} $h_{m,n}$ for all $m,n\in\Nodes$ as follows
\begin{equation}\label{eq:score}
 h_{m,n} = \sum_{k=1}^K h_{m,n}^{(k)}, 
 \quad\text{with}\quad
  h_{m,n}^{(k)}  
   = \begin{cases}
    \tfrac{1}{|\Infected_k|}, & \text{if}\: x_{n,k-1}=0, \: x_{n,k}=1, \:\text{and}\: m\in\Infected_k; \\
    0,                        & \text{otherwise},
     \end{cases}
\end{equation}
where 
\(
\Infected_k = \{ m\in\Nodes: x_{m,k-1}=1 \:\text{or}\: x_{m,k}=1 \}
\)
is the set of all infected nodes at the beginning or by the end of the interval $t\in[t_{k-1},t_k].$
The higher is the acquired score $h_{m,n}$ the higher is the evidence that $m\sim n$ in the contact network.
Hence, when the scores are calculated, we can sample an initial guess network $\Grid_0$ randomly with probabilities for each link proportional to the scores.
Alternatively, for a more deterministic approach, we can discretise the distribution, and set $m\sim n$ in $\Grid_0$ for all links $(m,n)$ for which the score exceeds the average,
\(
h_{m,n} \geq \frac{2}{N(N-1)} \sum_{i>j} h_{i,j}.
\)

\section{Results} \label{sec:results}           

The numerical experiments were implemented in Matlab 2022b based on TT-Toolbox\footnote{https://github.com/oseledets/TT-Toolbox} and tAMEn\footnote{https://github.com/dolgov/tamen} packages,
and run on one node of the HC44 series of the University of Bath ``Nimbus'' Microsoft Azure cluster.  
Experiments with different datasets were run in parallel over $42$ cores of the Intel Xeon Platinum 8168 CPUs.
Each of these parallel processes ran in a single--threaded mode.
The codes are made public and freely available from
 \href{https://github.com/savostyanov/ttsir}{github.com/savostyanov/ttsir}.

\subsection{Linear chain}  \label{sec:chain}    

\begin{figure}[tp]
   \includegraphics[width=\textwidth]{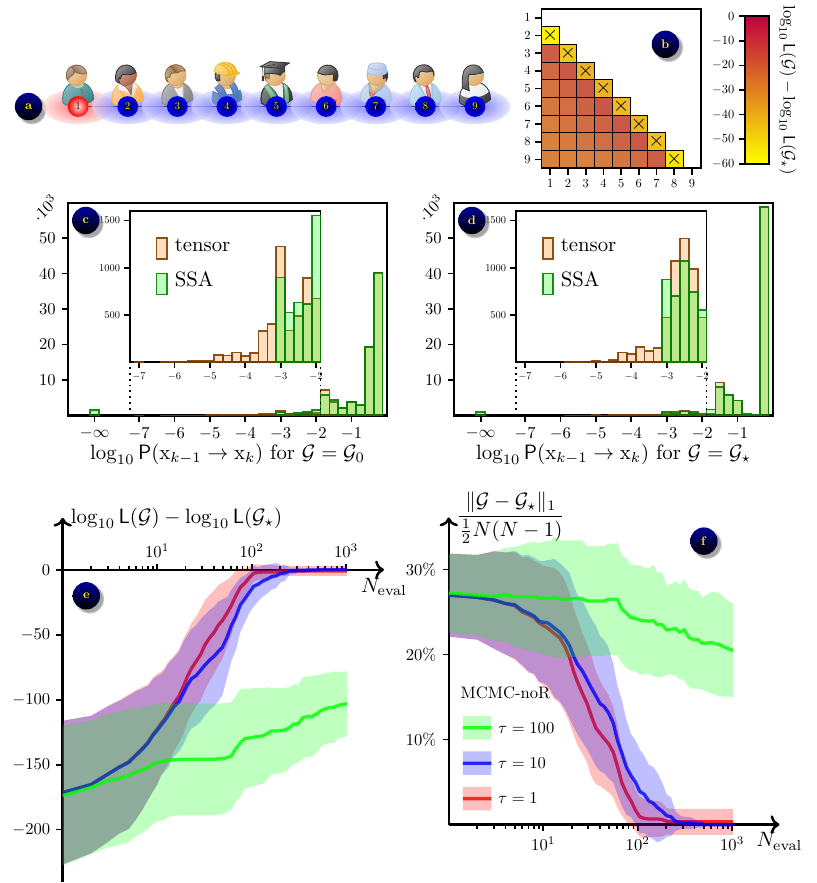}
 \caption{Inferring linear chain network with $N=9$ people from $\eps$--SIS epidemic process with $\beta=1,$ $\gamma=0.5$ and $\eps=0.01$:
            (a) the ground truth network $\Grid_\star$ in its initial state;
            (b) the contrast $\log_{10}\Like(\Grid)-\log_{10}\Like(\Grid_\star)$ averaged over $N_s=42$ datasets, shown for grids $\Grid$ that differ from $\Grid_\star$ by a single link $(m,n)$; axes $\times$ show links in $\Grid_\star$;
            (c) the distribution of probabilities for the transitions observed in data for the initial guess network $\Grid_0$;
            (d) the distribution of probabilities for the transitions observed in data for the ground truth network $\Grid_\star$;
            (e) convergence of likelihood $\Like(\Grid)$ towards $\Like(\Grid_\star)$ in the optimisation algorithm MCMC-noR; average (solid lines) $\pm$ one standard deviation (shaded areas) over the $N_s=42$ datasets; shown for temperatures $\Temp=1, 10, 100$;
            (f) convergence of network $\Grid$ towards $\Grid_\star.$
                 }
  \label{fig:chain}
\end{figure}

For this experiment we generated $N_s=42$ samples of synthetic data by computing random walks of $\eps$-SIS process with parameters $\beta=1,$ $\gamma=0.5$ and $\eps=0.01$ for the duration of $\tmax=200$ time units.
The trajectories were then re-sampled to a uniform grid on $[0,\tmax]$ with the time step $\tstep=0.1$ to imitate data collected at regular intervals.
Therefore, each trajectory contained $K=2000$ data records representing the epidemic process.
Data were created using the `ground truth' network $\Grid_\star$ which is a linear chain with $N=9$ nodes as shown in Fig.~\ref{fig:chain}(a), and assuming that the initial state $\x_0=(1,0,\ldots,0)$ is the same for all data samples.

First, we checked the contrast of the log--likelihood at the ground truth network $\Grid_\star,$ by computing
\(
\Mean{\log_{10}\Like(\Grid)-\log_{10}\Like(\Grid_\star)}
\)
for all $\Grid$ that are nearest neighbours of $\Grid_\star,$ i.e. differ by only one link.
The results are averaged over the $N_s=42$ sampled datasets and shown in Fig.~\ref{fig:chain}(b).
We note that removal of an existing link from $\Grid_\star$ results in contrast
\(
\Mean{\log_{10}\Like(\Grid)-\log_{10}\Like(\Grid_\star)} \simeq -50,
\)
raising to $\simeq-60$ for links attached to the sides of the chain.
This is easy to understand, as removal of a link from $\Grid_\star$ creates a disconnected network $\Grid,$ where two parts can not pass the infection on to each other, hence the epidemic dynamics on $\Grid$ differs significantly from the one on $\Grid_\star.$
Adding a new link to $\Grid_\star$ results in a milder contrast
\(
\Mean{\log_{10}\Like(\Grid)-\log_{10}\Like(\Grid_\star)} \in [-30,-10],
\)
because the grid remains connected and the dynamics of the epidemic is less affected.
This confirms that $\Grid_\star$ is at least a local optimum for $\log\Like(\Grid),$ and therefore can be inferred by Bayesian optimisation, assuming the optimisation algorithm manages to converge to it.

Secondly, we evaluated probabilities 
\(
\Prob(\x_{k-1}^{(n_s)}\to\x_k^{(n_s)} | \Grid)
\)
in~\eqref{eq:loglike} 
  for all data records $k=1,\ldots,K$ 
  for all generated datasets~$n_s=1,\ldots,N_s.$ 
The results are shown 
  in Fig.~\ref{fig:chain}(c) for the initial guess network $\Grid=\Grid_0$ computed as explained in Sec.~\ref{sec:init}, and
  in Fig.~\ref{fig:chain}(d) for the ground truth network $\Grid=\Grid_\star.$
We used the SSA algorithm~\cite{gillespie-ssa-1976} with $\Nssa=10^3$ samples as explained in Sec.~\ref{sec:SSA}.
A significant number of events are not resolved by SSA and the probabilities are estimated as zero, as shown by the $\log_{10}p=-\infty$ column on the histograms.
We then computed the same probabilities by solving the CME~\eqref{eq:CME} subject to initial condition $\x_{k-1}^{(n_s)}$ on time interval $t\in[t_{k-1},t_k],$ for which we apply the tAMEn algorithm~\cite{d-tamen-2018} with Chebysh\"ev polynomials of degree $12$ in time and relative accuracy threshold $\epsilon_{\text{tAMEn}}=10^{-6}.$
From the tAMEn algorithm we obtain the whole p.d.f. 
\(
\p(t) = \left[ p(\x,t) \right]_{\x\in\X^N}
\)
for \( t\in[t_{k-1},t_k] \) and for all states \( \x\in\X^N, \)
from which we extract the required probability by projecting to the deterministic final state \( \x_k^{(n_s)}. \)

We observe that $1.7\%$ of probabilities are unresolved by SSA for $\Grid=\Grid_0$ and $1.1\%$ of probabilities are unresolved for $\Grid=\Grid_\star,$ which is nevertheless sufficient for both likelihoods $\Like(\Grid_0)=0$ and $\Like(\Grid_\star)=0$ to be unresolved for $100\%$ of data samples $n_s=1,\ldots,N_s.$

The number of SSA trajectories is set to approximately match the computational time of SSA and tensor product algorithms for the forward problem.
With tAMEn, the trajectories $\p(t)$ are computed in the TT format~\eqref{eq:tt} for which the TT ranks are determined adaptively.
For this example we observe TT ranks 
 $r \simeq 14.8 \pm 2.3$ for $\Grid=\Grid_0$ and 
 $r \simeq 11.1 \pm 0.9$ for $\Grid=\Grid_\star$
leading to computational time for the likelihood $\Like(\Grid)$ to be
 $\text{CPU time} \simeq 98 \pm 6.9$ seconds for $\Grid=\Grid_0$ and
 $\text{CPU time} \simeq 80 \pm 3.2$ seconds for $\Grid=\Grid_\star.$
 With the SSA algorithm, one likelihood computation took
 $\text{CPU time} \simeq 199 \pm 23.5$ seconds for $\Grid=\Grid_0$ and
 $\text{CPU time} \simeq 107 \pm 6.3$ seconds for $\Grid=\Grid_\star.$
Note that the forward problems become easier to solve as the optimisation process approaches the ground truth network both because the linear geometry of the chain matches the structure of the TT format, and because the easier reaction network admits larger time steps in SSA.
Due to the simplicity of the linear structure, a linear chain is an attractive model for study in quantum physics, see e.g a recent paper on the SIS model on a linear chain~\cite{merbis-ttsis-2023}.

We performed the black--box Bayesian optimisation using $\neval=10^3$ steps of the MCMC algorithms with and without replacement as explained in Sec.~\ref{sec:MCMC}. 
We observed similar performance of both algorithms, hence only the results for MCMC-noR are shown
 in Fig.~\ref{fig:chain}(e) for the convergence of the likelihood  $\Like(\Grid)$ towards the one of the ground truth network, $\Like(\Grid_\star),$ and
 in Fig.~\ref{fig:chain}(f) for the corresponding convergence of the network  $\Grid$ towards the ground truth network $\Grid_\star.$
The latter is measured using the number of incorrectly inferred links,
\begin{equation}\label{eq:norm}
 \| \Grid - \Grid_\star \|_1 = \left| \{ m,n\in\Nodes, \, m>n : g_{m,n} \neq g_{m,n}^\star \} \right|,
\end{equation}
related to the total number of possible links, $\tfrac12 N(N-1).$
 For both MCMC-R and MCMC-noR without tempering (with $\Temp=1$), we observe a steady convergence towards optimum with the ground truth grid correctly inferred in $40$ out of $42$ experiments and one link inferred incorrectly in $2$ out of $42$ experiments after $\neval=10^3$ likelihood evaluations.
 To improve this result, we used tempering with temperature $\Temp=10,$ and observed a slightly slower convergence of MCMC-noR, which then achieved the exact recovery for all $N_s$ data samples.
 We also observed that increasing the temperature further to $\Temp=100$ results in a much slower convergence and poor recovery, indicating that this parameter needs to be carefully adjusted.
In total for this experiment, the network inference from each dataset with $K=2000$ records took about $7.3 \cdot 10^3$ seconds.

\subsection{Austria road network} \label{sec:austria} 
\begin{figure}[tp]
   \includegraphics[width=\textwidth]{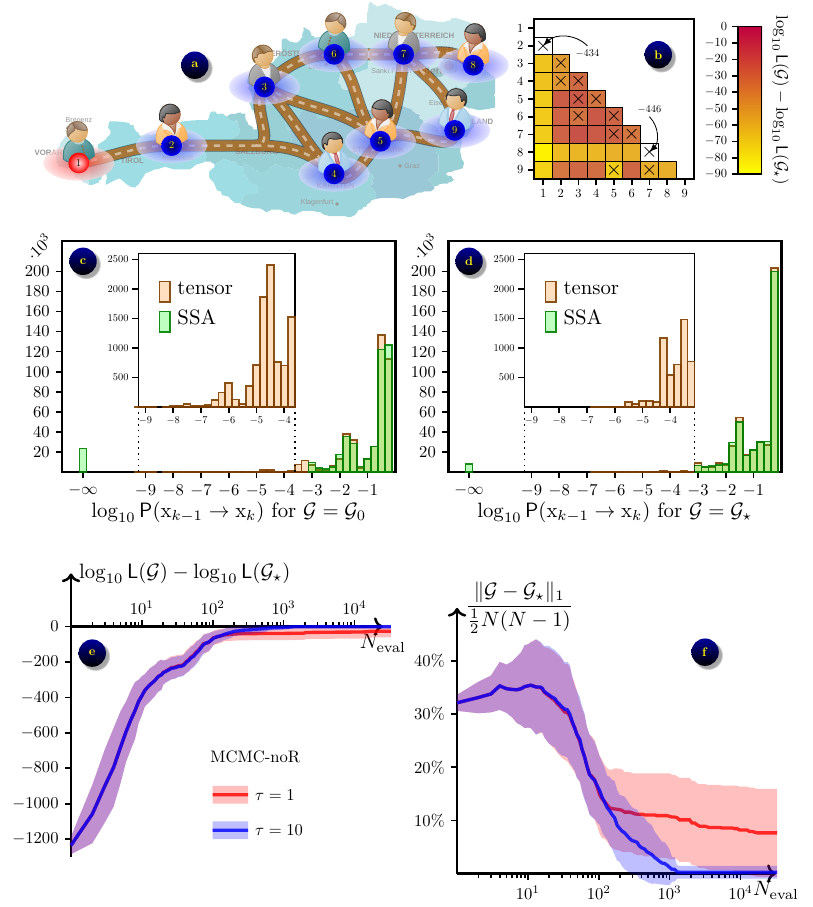}
 \caption{Inferring a road network in Austria ($N=9$ nodes) from $\eps$--SIS epidemic process with $\beta=1,$ $\gamma=0.5$ and $\eps=0.01$:
            (a) the ground truth network $\Grid_\star$ in its initial state;
            (b) the contrast $\log_{10}\Like(\Grid)-\log_{10}\Like(\Grid_\star)$ averaged over $N_s=42$ datasets, shown for grids $\Grid$ that differ from $\Grid_\star$ by a single link $(m,n)$; axes $\times$ show links in $\Grid_\star$;
            (c) the distribution of probabilities for the transitions observed in data for the initial guess network $\Grid_0$;
            (d) the distribution of probabilities for the transitions observed in data for the ground truth network $\Grid_\star$;
            (e) convergence of likelihood $\Like(\Grid)$ towards $\Like(\Grid_\star)$ in the optimisation algorithm MCMC-noR; average (solid lines) $\pm$ one standard deviation (shaded areas) over the $N_s=42$ datasets; shown for temperatures $\Temp=1, 10$;
            (f) convergence of network $\Grid$ towards $\Grid_\star.$
         }
  \label{fig:austria}
\end{figure}

For this experiment we considered a more realistic example of a contact network, drawn from the road network in Austria, shown in Fig.~\ref{fig:austria}(a).
As previously,
we generated $N_s=42$ samples of synthetic data for a $\eps$--SIS model with per contact transfer rate $\beta=1,$ individual recovery rate $\gamma=0.5$ and self--infection rate $\eps=0.01.$
However, from preliminary experiments we noted that both MCMC algorithms for Bayesian optimisation struggle to converge to the optimum.
To partly mitigate this, we increased the size of each dataset
to $K=10^4$ data records,
created by observing the state
for the duration of $\tmax=1000$ time units at uniform time grid with the step $\tstep=0.1.$

From the contrasts shown on Fig.~\ref{fig:austria}(b), we see that removal of any of two links that produces a disconnected graph $\Grid$ results in a very high contrast,
\(
\Mean{\log_{10}\Like(\Grid)-\log_{10}\Like(\Grid_\star)} \lesssim -400.
\)
Removing or adding other links results in a connected $\Grid$ and hence a moderate value of the contrast
\(
\Mean{\log_{10}\Like(\Grid)-\log_{10}\Like(\Grid_\star)} \in [-100,-15].
\)

Similarly to the previous example, we observe that SSA with $\nssa=10^3$ samples does not resolve a significant number of events along the trajectory, and therefore returns
\(
\tilde \Prob(\x_{k-1}\to\x_k | \Grid) = 0
\)
for $5.6\%$ of data points for the initial guess network $\Grid=\Grid_0$ and 
for $2.0\%$ of data points for the ground truth network $\Grid=\Grid_\star,$
leading to the likelihood $\Like(\Grid)=0$ being unresolved in all experiments for both grids.
Note that the proportion of unresolved (rare) events is larger for this example due to a more complex network structure.

Using the tensor product approach with the same parameters as in Sec.~\ref{sec:chain} for the forward problem, we were able to resolve probabilities of up to $p\sim 10^{-7},$ which produced non--zero values for all likelihoods $\Like(\Grid)$, enabling the optimisation for the inverse problem.
For this example, one likelihood evaluation solving the forward problem with tAMEn took
 $\text{CPU time} \simeq 431 \pm 8.2$ seconds for $\Grid=\Grid_0$ and
 $\text{CPU time} \simeq 439 \pm 4.8$ seconds for $\Grid=\Grid_\star.$
The main reason for the larger times compared to the previous experiment is the larger data size $K=10^4$ compared to $K=2000$ for the linear chain.
However, a more complex structure of the contact network also contributed via higher TT ranks
 $r \simeq 12.0 \pm 1.8$ for $\Grid=\Grid_0$ and
 $r \simeq 13.4 \pm 1.2$ for $\Grid=\Grid_\star.$
The total time required to perform the Bayesian optimisation with $\neval=10^4$ steps of the MCMC algorithm took us
about $45$ hours.
For comparison, when SSA is used as the forward solver, one likelihood computation took
 $\text{CPU time} \simeq 1292 \pm 58.7$ seconds for $\Grid=\Grid_0$ and
 $\text{CPU time} \simeq 853 \pm 24.4$ seconds for $\Grid=\Grid_\star.$

From results shown in Fig.~\ref{fig:austria} we note that without tempering ($\Temp=1$) the convergence of both MCMC algorithms is stuck in a local maximum where approximately $4$ of $36$ links are inferred incorrectly.
 Using tempering with $\Temp=10,$ we observed almost the same convergence at initial stage of optimisation, which then resulted in a faster convergence towards a better inference, with $38$ out of $42$ data samples allowed for the exact recovery, and the remaining $4$ out of $42$ had only one incorrect link out of $36.$

The difference in performance compared to the linear chain example can be considered as a consequence of a high contrast, that sharpens the high--dimensional landscape and makes both the global and local maxima steeper.
If we use less data for Bayesian inference, the contrast reduces, making it easier for MCMC to escape from local optima by switching from current network $\Grid_i$ to less attractive proposal $\hat\Grid$ with probability 
\(
\Like(\hat\Grid)/\Like(\Grid_i) < 1
\)
as explained in Alg.~\ref{alg:mcmc}.
However, it also makes global maximum less emphasised and can lead to a situation where the optimal grid recovered by the Bayesian optimisation is not the same as the ground truth grid, 
\(
\Grid_{\text{opt}} \neq \Grid_\star.
\)

\subsection{Florentine families} \label{sec:florence}
\begin{figure}[ht]
   \includegraphics[width=\textwidth]{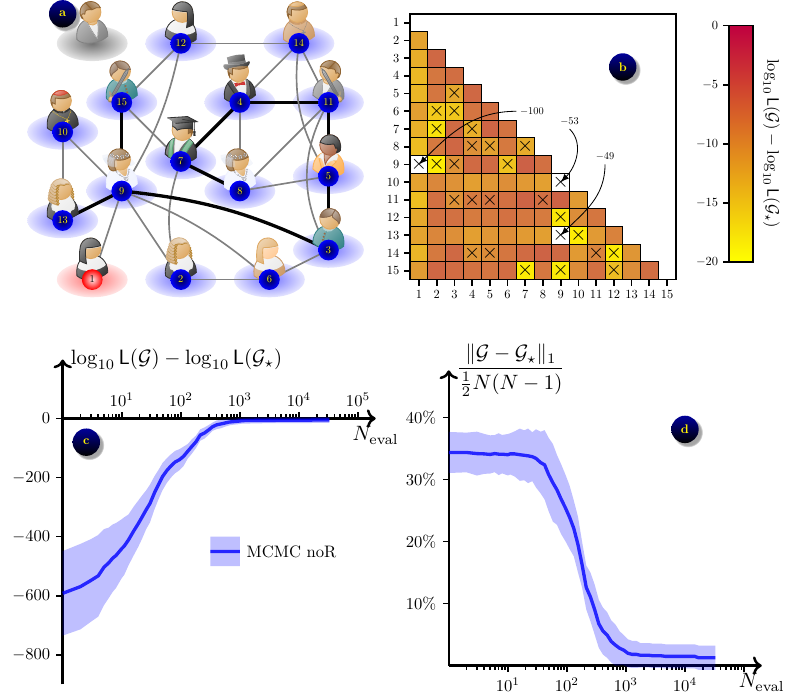}
 \caption{Inferring a network of Florentine families ($N=15$ nodes) from $\eps$--SIS epidemic process with $\beta=0.4,$ $\gamma=0.5$ and $\eps=0.004$:
            (a) the ground truth network $\Grid_\star$ in its initial state;
            (b) the contrast $\log_{10}\Like(\Grid)-\log_{10}\Like(\Grid_\star)$ averaged over $N_s=21$ datasets, shown for grids $\Grid$ that differ from $\Grid_\star$ by a single link $(m,n)$; axes $\times$ show links in $\Grid_\star$;
            (c) convergence of likelihood $\Like(\Grid)$ towards $\Like(\Grid_\star)$ in the optimisation process; average (solid lines) $\pm$ one standard deviation (shaded areas) over the $N_s=21$ datasets;
            (d) convergence of network $\Grid$ towards $\Grid_\star.$
         }
  \label{fig:florence}
\end{figure}

We consider a slightly larger network representing marriage alliances and business relationships between Florentine families in XV century.\footnote{The network is taken from \href{https://networks.skewed.de/net/florentine_families}{networks.skewed.de/net/florentine\_families}.}
The network is given as an undirected weighted graph with $16$ nodes, one of which is isolated from the rest, as shown in Figure~\ref{fig:florence}(a).
The weights of the links represent what kind of relationship the families have. 
For the purpose of this experiment we ignore the disconnected node and disregard the difference in connections.
Hence we consider an undirected and unweighted network of $N=15$ nodes.
Still, a more densely connected network leaves states more frequently in the infected state with $\beta=1$ used in the previous examples.
To obtain a more meaningful data (and more accurate inference), 
   we simulate the observation data using $\beta=0.4$, $\gamma=0.5$, and $\eps=0.004$.
We observe states at uniformly distributed time points 
   over the duration of $\tmax=400$ time units,
   sampled with the time step $\tstep=0.1,$
   resulting in $K=4\cdot10^3$ data records.

From the contrasts shown in Figure~\ref{fig:florence}(b) we see that removal of the link between nodes $1$ and $9$ results in disconnected network, which results in a significant contrast,
 \(
 \Mean{\log_{10}\Like(\Grid) - \log_{10}\Like(\Grid_\star)} \lesssim -100.
 \)
Notably, removal of the link $(9,10)$ does not fully disconnect the network, but considerably reduces the chance for the disease to reach from the node $9$ to the node $13,$ hence the observed large contrast 
 \(
 \Mean{\log_{10}\Like(\Grid) - \log_{10}\Like(\Grid_\star)} \approx -53.
 \)
 Similarly, removal of the link $(9,13)$ makes it harder for the disease to reach the node $10,$ which also results in a high contrast
 \(
 \Mean{\log_{10}\Like(\Grid) - \log_{10}\Like(\Grid_\star)} \approx -49.
 \)
 The remaining links seem to be considerably less important, and their removal results in a lower contrast
 \(
 \Mean{\log_{10}\Like(\Grid) - \log_{10}\Like(\Grid_\star)} \gtrsim -20.
 \)
 On average, adding extra links also result in a lower contrast, with a notable exception of the first node, which is the origin of the epidemic.
The lower contrast around the ground truth network may cause extra challenge for the exact recovery of this network, particularly if the likelihoods are computed inaccurately.

 We run the MCMC-noR algorithm without tempering ($\Temp=1$),
  and with tempering ($\Temp=10$),
  but observe better results of the former.
The convergence of the log-likelihood is shown in Figure~\ref{fig:florence}(c),
and the convergence of the inferred network is shown in Figure~\ref{fig:florence}(d).
We observe accurate recovery of the network, specifically,
  among the $N_s=21$ data samples that we tried, $11$ resulted in exact recovery.
In the remaining $10$ data samples, we observed only $1$ to $3$ incorrectly recovered links.
In our experiments, the MCMC-noR algorithm has reached the final network configuration after $\neval \approx 41 \cdot 10^3$ samples on average, which took $4.2$ days of CPU time.

\subsection{Small world network} \label{sec:sw} 

\begin{figure}[t]
   \includegraphics[width=\textwidth]{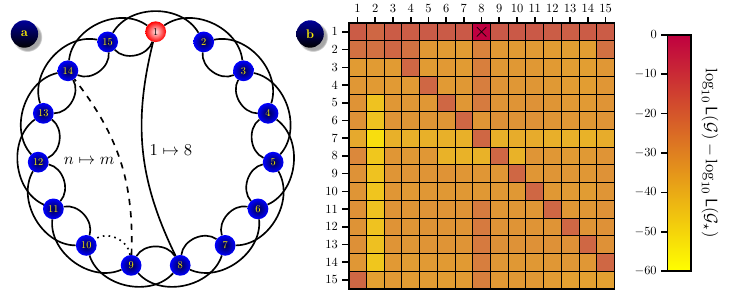}
 \caption{Inferring a rewired link in small world graph ($N=15$ nodes) from $\eps$--SIS epidemic process with $\beta=1,$ $\gamma=0.5$ and $\eps=0.01$:
         (a) the ground truth grid $\Grid_\star$ shown in its initial state;
         (b) the contrast $\log_{10}\Like(\Grid)-\log_{10}\Like(\Grid_\star)$ averaged over $N_s=42$ datasets, shown for grids $\Grid\in\tilde\Grids$ from a class of small--world networks with a single rewired link.
         }
  \label{fig:sw}
\end{figure}

We note that even though the use of tensor product algorithms allows us to compute likelihoods~\eqref{eq:loglike} faster and more accurately, the exact Bayesian inference of a contact network in a fully black--box setting remains a challenging task, as we see from experiments in Section~\ref{sec:austria} and Section~\ref{sec:florence}.

In this section we present a preliminary experiment where we assume some prior knowledge of the contact network, which allows us to reduce the number of unknown parameters even for a larger number of network nodes.
Specifically, we assume that the contact network is from a family of small--world networks~\cite{strogatz-sw-1998}, which is shown in Fig.~\ref{fig:sw}(a).
It consists of $N=15$ nodes which are arranged as a loop and connected with a double bond, where each node $n\in\Nodes$ is connected to nodes $n+1$ and $n+2,$ where we assume that indices go around the circle, so $N+1=1$ and $N+2=2$ when necessary.
The main loop is rewired, i.e a certain link $(n,n+1)$ is removed and replaced with a link $(n,m)$ to a random node $m\in\Nodes,$ which provides additional connectivity.
For this experiment we assumed that the ground truth network $\Grid_\star$ contains a single rewired link $1\mapsto8,$ i.e. the link $(1,2)$ is removed and replaced with $(1,8).$
We then proceed to infer this network, assuming that we know it is from the set of small--world networks with a single rewired link $n\mapsto m,$ which we denote $\tilde\Grids.$ 
The problem therefore reduces to finding only two parameters, $n$ and $m,$ and the search space shrinks from 
$|\Grids|=2^{N(N-1)/2}$ to only $|\tilde\Grids|=N^2$ possible grids.

Inferring a network from a known class can be formulated as Bayesian optimisation~\eqref{eq:loglike} on a class of networks $\tilde\Grids$ parameterised by a small number of parameters.
This removes our main computational challenge related to high dimensionality of the search space and allows us to solve this problem directly.
We generated  $N_s=42$ data samples by simulating the $\eps$--SIS epidemic on a ground truth contact network $\Grid_\star$ using parameters $\beta=1,$ $\gamma=0.5$ and $\eps=0.01,$ for the duration of $\tmax=1000$ time units, and re-sampled the data to a uniform grid with the time step $\tstep=0.1,$ hence creating $K=10^4$ records for each data sample.
Using tAMEn algorithm to model the evolution of epidemic on $15$--node networks, we were able to compute the likelihoods for all grids $\Grid\in\tilde\Grids.$
We then computed the average contrast for all $\Grid\in\tilde\Grids$ as shown in Fig.~\ref{fig:sw}(b).
The results show that
\(
\Mean{\log_{10}\Like(\Grid)-\log_{10}\Like(\Grid_\star)} \leq -10
\)
for all $\Grid\neq\Grid_\star,$
which ensures that the ground truth network is a unique global maximum of the Bayesian optimisation problem~\eqref{eq:loglike}.

\section{Discussion}  \label{sec:discussion} 

Inferring the contact network in a Bayesian optimisation framework requires us to estimate the likelihood of observed data $\Data,$ which are a realisation of epidemic dynamics on the ground truth network $\Grid_\star,$ to appear for the epidemic on another network $\Grid.$
In a black--box setting, we have no a priori information on the network, and start the optimisation from an initial guess $\Grid_0$ that may be (very) different from $\Grid_\star.$
For the grids $\Grid$ in the vicinity of $\Grid_0,$ observing the same dynamics as on $\Grid_\star$ is a (very) rare event, which we need to estimate with sufficient precision in order to evaluate the likelihoods $\Like(\Grid).$ 
The slow convergence of the SSA algorithm limits its capability to recover rare events.
By replacing it with the tensor product algorithms, we are able to recover rare events much more accurately by solving the forward problem in the CME form~\eqref{eq:CME} and overcoming the curse of dimensionality.
This allows the MCMC method to find its way from the initial network $\Grid_0$ towards the optimum.

As the optimisation gets closer to $\Grid_\star,$ the likelihoods increase and the presence of steep local maxima slows down the convergence towards the global one.
In this area high contrast ratios $\Like(\Grid_\star)/\Like(\Grid)$ are undesirable as they make it harder for the MCMC algorithm to escape local maxima.
By preliminary experiments demonstrated in this paper we show that this can be addressed  tempering of $\Like(\Grid)$ to simplify the high--dimensional landscape for the optimisation.
Another idea is to use only a part of the available data to compute likelihoods~\eqref{eq:loglike}, which has been used successfully for sampling from concentrated distributions of continuous random variables~\cite{cd-DIRT-2022}.

We also explored the potential of tensor product algorithms for tackling the network likelihood optimisation.
However, these attempts so far were less efficient than the MCMC algorithm (in particular the MCMC without replacement).
The TT-Cross algorithm~\cite{ot-ttcross-2010} and its extensions~\cite{ds-parcross-2020} are used to compute a TT approximation to a black-box tensor by drawing a few adaptive samples from it using the maximum volume algorithm~\cite{gostz-maxvol-2010} or a greedy version thereof~\cite{ds-parcross-2020}.
These maximum volume samples are expected to be good candidates for the maximum absolute value of the tensor~\cite{gostz-maxvol-2010,scspco-ttopt-2022}.
However, the maximum volume algorithm requires all elements of a TT core, which must be drawn as full columns from the tensor, including elements which are known to be far from the maximum.
MCMC probes only one element at a time, and can skip such unnecessary calculations.
In numerical experiments with the linear chain, MCMC was systematically faster and more accurate compared to the TT-Cross maximiser, albeit by a modest margin (1-2 contacts).
Tempering the likelihood to reduce its TT ranks and caching its values (which are often repeated in the TT-Cross) may make this approach faster in terms of the actual CPU time.

Another tensor optimiser proposed recently is PROTES \cite{bcro-protes-2023}, a probabilistic method similar to genetic algorithms.
In each iteration, this algorithm draws $N_c$ candidate optima as random samples from a probability distribution function in the TT format, which is in turn updated by a stochastic gradient ascent maximising the probability of drawing $n_s$ samples with the largest values of the sought function out of the $N_c$ candidates.
The default parameters proposed in \cite{bcro-protes-2023} are $n_s=10$ and $N_c=100$.
Compared to our budget of $\neval=400$ function evaluations, this corresponds to only $4$ stochastic gradient ascent iterations, which are clearly insufficient and produce an almost random network.
Taking $N_c$ in the order of $10$ (and hence $n_s < 10$) is uncompetitive too, since a few tens of iterations cannot compensate for a more random stochastic gradient due to a smaller $n_s$.
However, it may be reasonable to use such an algorithm to fine-tune a previous TT approximation of the likelihood to new data.

Choosing a more informative prior on the network may aid the inference.
We have already stepped away from a fully uniform prior in the small world example, where we sought only one rewiring instead of the state of all links.
Penalising improbable or redundant links with a low prior probability may be beneficial for more general networks as well.

Potentially, it may be possible to use all MCMC points to compute posterior expectations rather than the MLE/MAP.
However, this would be difficult for network identification for two reasons.
First, accurate sampling would need much more likelihood evaluations (and hence CPU time) to decorrelate the Markov chain, whereas the MLE/MAP can be found in a few thousand samples.
Secondly, the expected state would be a real-valued instead of binary vector, and require ambiguous post-processing to convert it into a network.
Mitigation of these obstacles can be a matter of a future research.


\newpage
\backmatter
{\small
\section*{Abbreviations}
\begin{itemize}
 \item \textbf{SIR} = susceptible--infected--recovered epidemic model, see~\cite{kmk-sir-1927}
 \item \textbf{SIS} = susceptible--infected--susceptible epidemic model, see e.g.~\cite{mieghem-sis-2009} 
 \item \textbf{ODE} = ordinary differential equation
 \item \textbf{CME} = chemical master equation, see~\cite{vankampen-stochastic-1981}
 \item \textbf{MLE} = maximum likelihood estimate, see e.g.~\cite{bain-stats-1992}
 \item \textbf{MAP} = maximum a posteriori estimate, see e.g.~\cite{bain-stats-1992}
 \item \textbf{SSA} = stochastic simulation algorithm, see~\cite{gillespie-ssa-1976}
 \item \textbf{MCMC} = Markov chain Monte Carlo~\cite{RobertsRosenthal2011}
 \item \textbf{MCMC-noR} = a version of MCMC without replacement, see Sec.~\ref{sec:MCMC}
 \item \textbf{CP} = canonical polyadic tensor product format, see~\cite{jahnke-cme-2008,Ammar-cme-2011,hegland-cme-2011}
 \item \textbf{TT} = tensor train format, see~\cite{osel-tt-2011}
 \item \textbf{MPS} = matrix product states, see e.g.~\cite{schollwock-2011}
 \item \textbf{DMRG} = density matrix renormalisation group algorithm, see~\cite{white-dmrg-1993}
 \item \textbf{AMEn} = alternating minimal energy method, see~\cite{ds-amen-2014}
 \item \textbf{tAMEn} = time--dependent AMEn, see~\cite{d-tamen-2018}
 \item \textbf{CPU time} = central processing unit time, also known as the wallclock time
\end{itemize}

\section*{Declarations}

\bmhead{Funding}
Sergey Dolgov was supported by the Engineering and Physical Sciences Research Council (EPSRC) New Investigator Award EP/T031255/1.
Dmitry Savostyanov was supported by the Leverhulme Trust Research Fellowship RF-2021-258 at the initial stage of this work.
Dmitry Savostyanov would like to thank the Isaac Newton Institute for Mathematical Sciences, Cambridge, for support and hospitality during the programme Discretization and recovery in high-dimensional spaces, where work on the revised version of this paper was undertaken. This work was partly supported by EPSRC grant no EP/R014604/1 and by a grant from the Simons Foundation.
Funders took no part in study design; in the collection, analysis and interpretation of data; in the writing of the report; and in the decision to submit the article for publication.

\bmhead{Conflict of interest}
Authors declare no competing interests exist.

\bmhead{Ethics approval and consent to participate}
Not applicable. 

\bmhead{Consent for publication}
Not applicable.

\bmhead{Data and code availability}
Numerical experiments in this work are based on synthetic randomly generated datasets. 
All data and code required to reproduce experiments are available on
 \href{https://github.com/savostyanov/ttsir}{github.com/savostyanov/ttsir}.

\bmhead{Materials availability}
Not applicable.

\bmhead{Authors' contribution}
Sergey Dolgov developed software, performed numerical experiments, analysed the results, and contributed to writing the manuscript.
Dmitry Savostyanov designed the work, analysed the results, designed visualisations, and was a major contributor to writing the manuscript.
Both authors read and approved the final manuscript.
}

\newpage


\end{document}